\documentstyle[twoside,epsfig]{article}

\input{psfig.tex}

\newcommand{\D}{{\rm d}}
\newcommand{\abst}{\mbox{$|t|$}}
\newcommand{\gev}{{\rm Ge}\kern-1.pt{\rm V}}
\newcommand{\tev}{{\rm Te}\kern-1.pt{\rm V}}
\newcommand{\gevsq}{\mbox{$\mathrm{{\rm Ge}\kern-1.pt{\rm V}}^2$}}

\newcommand{\rhoz}{\mbox{$\rho^0$}}
\newcommand{\jpsi}{\mbox{$J/\psi$}}
\newcommand{\qsq}{\mbox{$Q^2$}}  

\catcode`\@=11
\long\def\@makefntext#1{
\protect\noindent \hbox to 3.2pt {\hskip-.9pt  
$^{{\eightrm\@thefnmark}}$\hfil}#1\hfill}               

\def\@makefnmark{\hbox to 0pt{$^{\@thefnmark}$\hss}}    
        
\def\ps@myheadings{\let\@mkboth\@gobbletwo
\def\@oddhead{\hbox{}
\rightmark\hfil\eightrm\thepage}   
\def\@oddfoot{}\def\@evenhead{\eightrm\thepage\hfil
\leftmark\hbox{}}\def\@evenfoot{}
\def\sectionmark##1{}\def\subsectionmark##1{}}

\oddsidemargin=\evensidemargin
\newcounter{sectionc}\newcounter{subsectionc}\newcounter{subsubsectionc}
\renewcommand{\section}[1] {\vspace{12pt}\addtocounter{sectionc}{1} 
\setcounter{subsectionc}{0}\setcounter{subsubsectionc}{0}\noindent 
        {\tenbf\thesectionc. #1}\par\vspace{5pt}}
\renewcommand{\subsection}[1] {\vspace{12pt}\addtocounter{subsectionc}{1} 
        \setcounter{subsubsectionc}{0}\noindent 
        {\bf\thesectionc.\thesubsectionc. {\kern1pt \bfit #1}}\par\vspace{5pt}}
\renewcommand{\subsubsection}[1] {\vspace{12pt}\addtocounter{subsubsectionc}{1}
        \noindent{\tenrm\thesectionc.\thesubsectionc.\thesubsubsectionc.
        {\kern1pt \tenit #1}}\par\vspace{5pt}}
\newcommand{\nonumsection}[1] {\vspace{12pt}\noindent{\tenbf #1}
        \par\vspace{5pt}}

\newcounter{appendixc}
\newcounter{subappendixc}[appendixc]
\newcounter{subsubappendixc}[subappendixc]
\renewcommand{\thesubappendixc}{\Alph{appendixc}.\arabic{subappendixc}}
\renewcommand{\thesubsubappendixc}
        {\Alph{appendixc}.\arabic{subappendixc}.\arabic{subsubappendixc}}

\renewcommand{\appendix}[1] {\vspace{12pt}
        \refstepcounter{appendixc}
        \setcounter{figure}{0}
        \setcounter{table}{0}
        \setcounter{lemma}{0}
        \setcounter{theorem}{0}
        \setcounter{corollary}{0}
        \setcounter{definition}{0}
        \setcounter{equation}{0}
        \renewcommand{\thefigure}{\Alph{appendixc}.\arabic{figure}}
        \renewcommand{\thetable}{\Alph{appendixc}.\arabic{table}}
        \renewcommand{\theappendixc}{\Alph{appendixc}}
        \renewcommand{\thelemma}{\Alph{appendixc}.\arabic{lemma}}
        \renewcommand{\thetheorem}{\Alph{appendixc}.\arabic{theorem}}
        \renewcommand{\thedefinition}{\Alph{appendixc}.\arabic{definition}}
        \renewcommand{\thecorollary}{\Alph{appendixc}.\arabic{corollary}}
        \renewcommand{\theequation}{\Alph{appendixc}.\arabic{equation}}
        \noindent{\tenbf Appendix \theappendixc #1}\par\vspace{5pt}}
\newcommand{\subappendix}[1] {\vspace{12pt}
        \refstepcounter{subappendixc}
        \noindent{\bf Appendix \thesubappendixc. {\kern1pt \bfit #1}}
        \par\vspace{5pt}}
\newcommand{\subsubappendix}[1] {\vspace{12pt}
        \refstepcounter{subsubappendixc}
        \noindent{\rm Appendix \thesubsubappendixc. {\kern1pt \tenit #1}}
        \par\vspace{5pt}}

\newcommand{\textlineskip}{\baselineskip=13pt}
\newcommand{\smalllineskip}{\baselineskip=10pt}

\def\eightcirc{
\begin{picture}(0,0)
\put(4.4,1.8){\circle{6.5}}
\end{picture}}
\def\eightcopyright{\eightcirc\kern2.7pt\hbox{\eightrm c}} 

\newcommand{\copyrightheading}[1]
        {\vspace*{-2.5cm}\smalllineskip{\flushright 
        {\footnotesize BONN-HE-2000-03}\\
        {\footnotesize October, 2000}\\
         }}

\def\abstracts#1#2#3{{
        \centering{\begin{minipage}{4.5in}\baselineskip=10pt\footnotesize
        \parindent=0pt #1\par 
        \parindent=15pt #2\par
        \parindent=15pt #3
        \end{minipage}}\par}} 

\newcommand{\bibit}{\nineit}

\renewenvironment{thebibliography}[1]
        {\frenchspacing
         \ninerm\baselineskip=11pt
         \begin{list}{\arabic{enumi}.}
        {\usecounter{enumi}\setlength{\parsep}{0pt}
         \setlength{\leftmargin 12.7pt}{\rightmargin 0pt} 
         \setlength{\itemsep}{0pt} \settowidth
        {\labelwidth}{#1.}\sloppy}}{\end{list}}

\newcounter{itemlistc}
\newcounter{romanlistc}
\newcounter{alphlistc}
\newcounter{arabiclistc}

\newcommand{\fcaption}[1]{
        \refstepcounter{figure}
        \setbox\@tempboxa = \hbox{\footnotesize Fig.~\thefigure. #1}
        \ifdim \wd\@tempboxa > 5in
           {\begin{center}
        \parbox{5in}{\footnotesize\smalllineskip Fig.~\thefigure. #1}
            \end{center}}
        \else
             {\begin{center}
             {\footnotesize Fig.~\thefigure. #1}
              \end{center}}
        \fi}

\newcommand{\tcaption}[1]{
        \refstepcounter{table}
        \setbox\@tempboxa = \hbox{\footnotesize Table~\thetable. #1}
        \ifdim \wd\@tempboxa > 5in
           {\begin{center}
        \parbox{5in}{\footnotesize\smalllineskip Table~\thetable. #1}
            \end{center}}
        \else
             {\begin{center}
             {\footnotesize Table~\thetable. #1}
              \end{center}}
        \fi}

\def\@citex[#1]#2{\if@filesw\immediate\write\@auxout
        {\string\citation{#2}}\fi
\def\@citea{}\@cite{\@for\@citeb:=#2\do
        {\@citea\def\@citea{,}\@ifundefined
        {b@\@citeb}{{\bf ?}\@warning
        {Citation `\@citeb' on page \thepage \space undefined}}
        {\csname b@\@citeb\endcsname}}}{#1}}

\newif\if@cghi
\def\cite{\@cghitrue\@ifnextchar [{\@tempswatrue
        \@citex}{\@tempswafalse\@citex[]}}
\def\citelow{\@cghifalse\@ifnextchar [{\@tempswatrue
        \@citex}{\@tempswafalse\@citex[]}}
\def\@cite#1#2{{$\null^{#1}$\if@tempswa\typeout
        {IJCGA warning: optional citation argument 
        ignored: `#2'} \fi}}

\def\pmb#1{\setbox0=\hbox{#1}
        \kern-.025em\copy0\kern-\wd0
        \kern.05em\copy0\kern-\wd0
        \kern-.025em\raise.0433em\box0}

\def\fnt#1#2{\footnotetext{\kern-.3em
        {$^{\mbox{\scriptsize #1}}$}{#2}}}

\def\fpage#1{\begingroup
\voffset=.3in
\thispagestyle{empty}\begin{table}[b]\centerline{\footnotesize #1}
        \end{table}\endgroup}

\def\runninghead#1#2{\pagestyle{myheadings}
\markboth{{\protect\footnotesize\it{\quad #1}}\hfill}
{\hfill{\protect\footnotesize\it{#2\quad}}}}
\font\tenrm=cmr10
\font\tenit=cmti10 
\font\tenbf=cmbx10
\font\bfit=cmbxti10 at 10pt
\font\ninerm=cmr9
\font\nineit=cmti9

\font\eightrm=cmr8

\def\qed{\hbox{${\vcenter{\vbox{                        
   \hrule height 0.4pt\hbox{\vrule width 0.4pt height 6pt
   \kern5pt\vrule width 0.4pt}\hrule height 0.4pt}}}$}}

\begin{document}

\runninghead{Measurements of Diffractive Vector-Meson Photoproduction $\ldots$} 
{Measurements of Diffractive Vector-Meson Photoproduction $\ldots$} 

\normalsize\textlineskip
\thispagestyle{empty}
\setcounter{page}{1}

\copyrightheading{}                     

\vspace*{0.88truein}

\fpage{1}
\centerline{\bf Measurements of Diffractive Vector-Meson Photoproduction}
\vspace*{0.035truein}
\centerline{\bf at High Momentum Transfer
from the ZEUS Experiment at HERA
}
\vspace*{0.15truein}
\centerline{\footnotesize James A. Crittenden
\footnote{Work supported by the Federal Ministry for Education
and Research of Germany}
\footnote{Present address: Deutsches Elektronen-Synchrotron, Notkestr. 85, 22607 Hamburg, Germany}
}
\vspace*{0.015truein}
\centerline{\footnotesize\it Physikalisches Institut, University of Bonn, Nu{\ss}allee 12}
\baselineskip=10pt
\centerline{\footnotesize\it 53115 Bonn, Germany}
\vspace*{10pt}
\abstracts{
We discuss recent preliminary results on the diffractive photoproduction of
{\rhoz}, $\phi$, and {\jpsi} mesons at high momentum transfer reported
by the ZEUS collaboration at HERA.  A special-purpose
calorimeter served to tag the quasi-real photons
($\qsq<0.01~\gevsq$) in the process 
$\gamma + p \rightarrow {\rm VM}+ Y$, where $Y$ represents a dissociated 
state of the proton. The resulting range in photon-proton center-of-mass energy
extends from 80 to 120~\gev.
The differential cross sections
$\frac{\D\sigma}{\D t}$ were
obtained in the region ${\abst}>1.2~\gevsq$, where $t$ denotes the squared momentum transfer to the proton. The measurements provide good sensitivity
to the observed power-law dependence $\frac{\D\sigma}{\D t} \propto {(-t)}^{-n}$.
Over the region in momentum transfer covered by the data, the power is found to be approximately $n\simeq 3$ for {\rhoz} and $\phi$ photoproduction, and approximately $n\simeq 2$ for {\jpsi} photoproduction.
}{}{}
\vspace*{1pt}
\textlineskip      
\section{Introduction}    
\vspace*{-0.5pt}
\noindent
The broad measurement program  of vector-meson photo- and electroproduction
presented by the ZEUS and H1 experiments operating at the 
electron-proton collider HERA since 1992 has proven to be a 
rich source of information on the strong interaction.
In the context of quantum chromodynamical descriptions of these processes, 
factorization scales given by the photon virtuality, the momentum transfer
to the proton, and the mass of the vector meson have been proposed.
In this talk, we discuss recent preliminary results\cite{1} 
presented by the ZEUS
collaboration on {\rhoz}, $\phi$, and {\jpsi} photoproduction for values
of the squared momentum transfer {\abst} greater than 1.2~\gevsq. 
The hardest scale in this process for the light vector mesons
is thus the momentum transfer,
and comparison to {\jpsi} photoproduction yields information on the r\^ole of
the vector-meson mass. The broad range in momentum transfer
covered by these results
provides good sensitivity to the observed 
power-law dependence in the differential cross sections $\frac{\D\sigma}{\D t}$.
We discuss the
scaling characteristics of the cross sections 
and summarize the salient features
of these results, including the production ratios and vector-meson
polarization measurements.

\vspace*{1pt}\textlineskip   
\section{Differential cross section $\frac{\D\sigma}{\D t}$}
\vspace*{-0.5pt}
\noindent
Differential cross sections $\frac{\D\sigma}{\D t}$ for {\rhoz},
$\phi$, and {\jpsi} mesons were measured using data recorded in 1996
and 1997 which correspond to a total integrated luminosity of
24~pb$^{-1}$.  The trigger conditions required the scattered positron
to be detected in a special-purpose tungsten/scintillator calorimeter
located 3~cm from the positron beam axis, 44~meters distant from the
nominal e$^+$p interaction point in the positron-beam flight
direction.  The position of this photoproduction tagger determined the
accepted range of energy lost by the positron to the photon which
interacted with the proton, thus restricting the range in photon-proton
center-of-mass energy, $W$, to $80<W<120~\gev$.  Since the transverse
momentum of the final-state positron was required to be
small 
by the acceptance of the tagger (${\qsq}<0.01~{\gevsq}$), 
the transverse momentum of the vector meson, $p_{\rm t}$, provided an accurate estimate of the square of the momentum
transferred to the proton via $t\simeq -p^2_{\rm t}$. The {\rhoz}, $\phi$,
and {\jpsi} meson samples were selected by fitting invariant-mass
spectra of track pairs with a sum of functions for signal and
background.  Offline data selection criteria included the
reconstruction of exactly two tracks from the interaction vertex and
rejected events with energy deposits in the rear and
barrel sections of the calorimeter which 
did not match the
extrapolation of either track. The selected events exhibit a
semi-exclusive topology with a gap between the dissociated nucleonic
system and the two tracks of at least two units of rapidity.  Trigger
requirements limited the range in $t$ to \mbox{${\abst}>1~\gevsq$},
where the proton-dissociative process dominates over elastic
production.  The dependence on the invariant mass of the dissociated
proton system, $M_Y$, was estimated by comparing forward energy
deposit distributions with calorimeter response simulations. The
differential cross sections {$\frac{\D \sigma}{\D t}$}
were calculated by integrating over $M_Y$
up to the limit given by $M_Y^2 = 100*{\abst}$ for fixed~$t$.

\begin{figure}[htbp]
\vspace*{2mm}
\begin{minipage}[t]{\textwidth}
\begin{center}
\epsfig{file=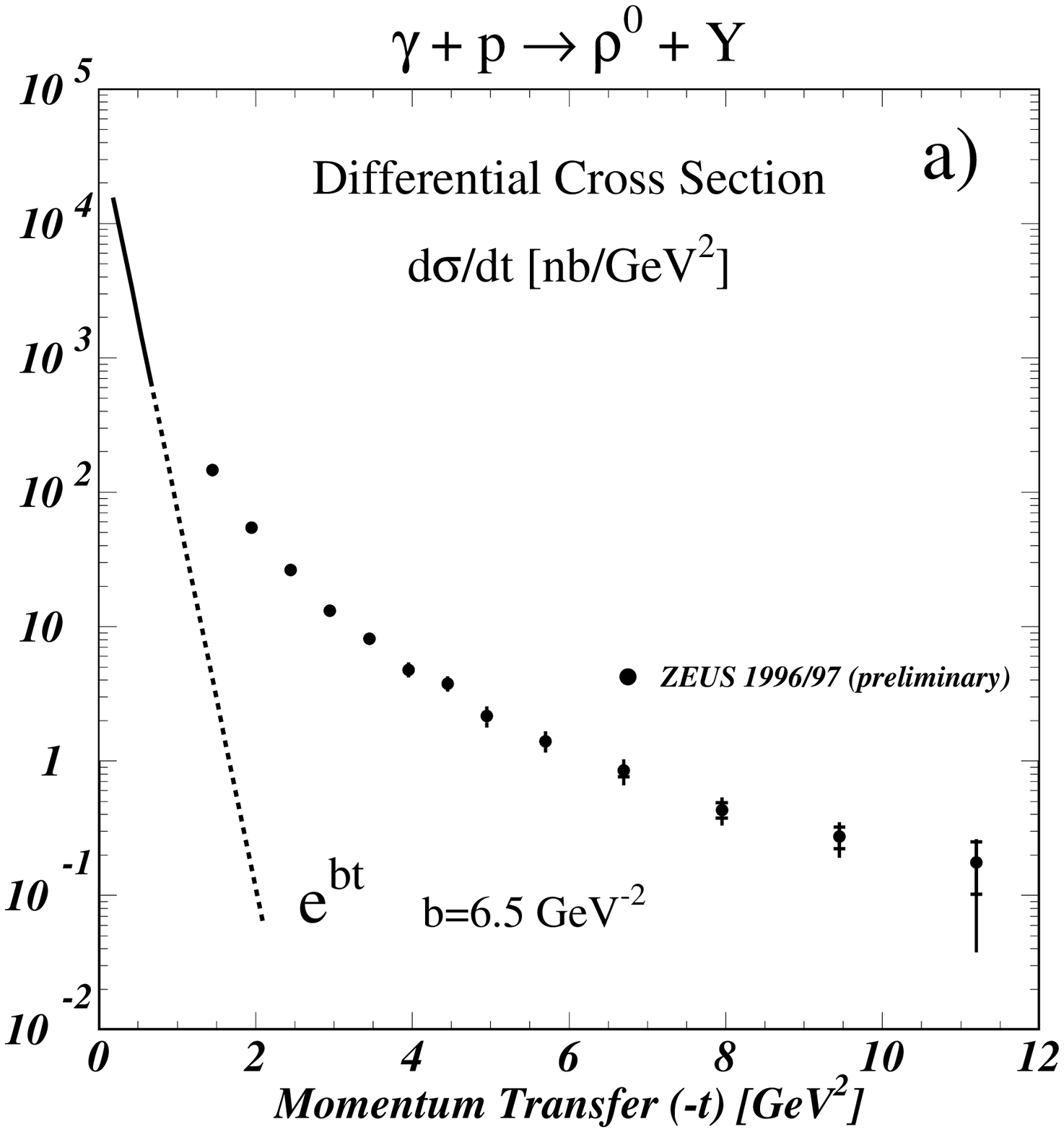, width=0.46\textwidth, bbllx=36, bblly=150, bburx=531, bbury=688, clip=}
\hfill
\epsfig{file=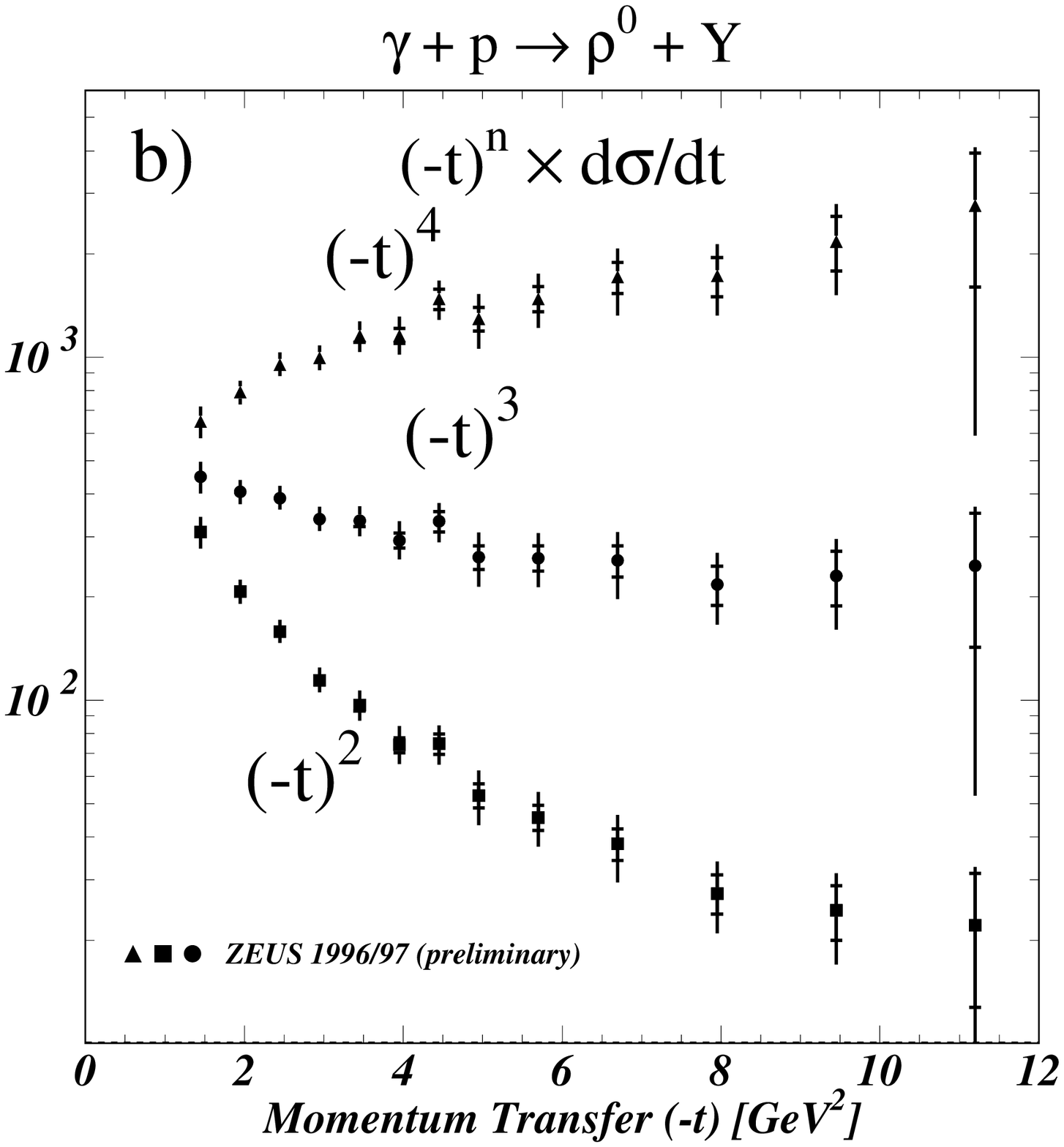, width=0.46\textwidth, bbllx=36, bblly=150, bburx=531, bbury=688, clip=}
\end{center}
\end{minipage}
\vspace*{5pt}
\fcaption{
\label{fig:rhoxsec}
a)~Differential cross section $\frac{\D\sigma}{\D t}$ for the process \mbox{$\gamma + p
\rightarrow {\rhoz} + Y$}, where $Y$ is a dissociated proton state. 
The line shows the $t$
dependence measured
for this process $\gamma + p \rightarrow {\rhoz} + Y$
($\propto e^{6.5\cdot t}$) at low {\abst} (solid line) and its extrapolation
to higher {\abst} (dashed line) for comparison. 
b)~Differential cross section $\frac{\D\sigma}{\D t}$  multiplied by {(-t)}$^n$, where $n$=2, 3, 4.
}
\end{figure}

Fig.~\ref{fig:rhoxsec}a compares the result for the {\rhoz}
cross section
to the extrapolation of the exponential behavior
observed\cite{2} at low {\abst}, showing that it underestimates the cross
section by an order of magnitude even at the lowest {\abst} value of the 
new result.

Fig.~\ref{fig:rhoxsec}b investigates the validity of the hypothesis of
a power-law dependence by scaling the cross section values with the
function $(-t)^{n}$, for \mbox{$n=2,3,4$}. While $n=3$ describes the
data best over the measured region, there is evidence for a
contribution steeper than a simple power law for ${\abst}<4~\gevsq$.

\begin{figure}[htbp]
\vspace*{-1mm}
\begin{minipage}[t]{\textwidth}
\begin{center}
\epsfig{file=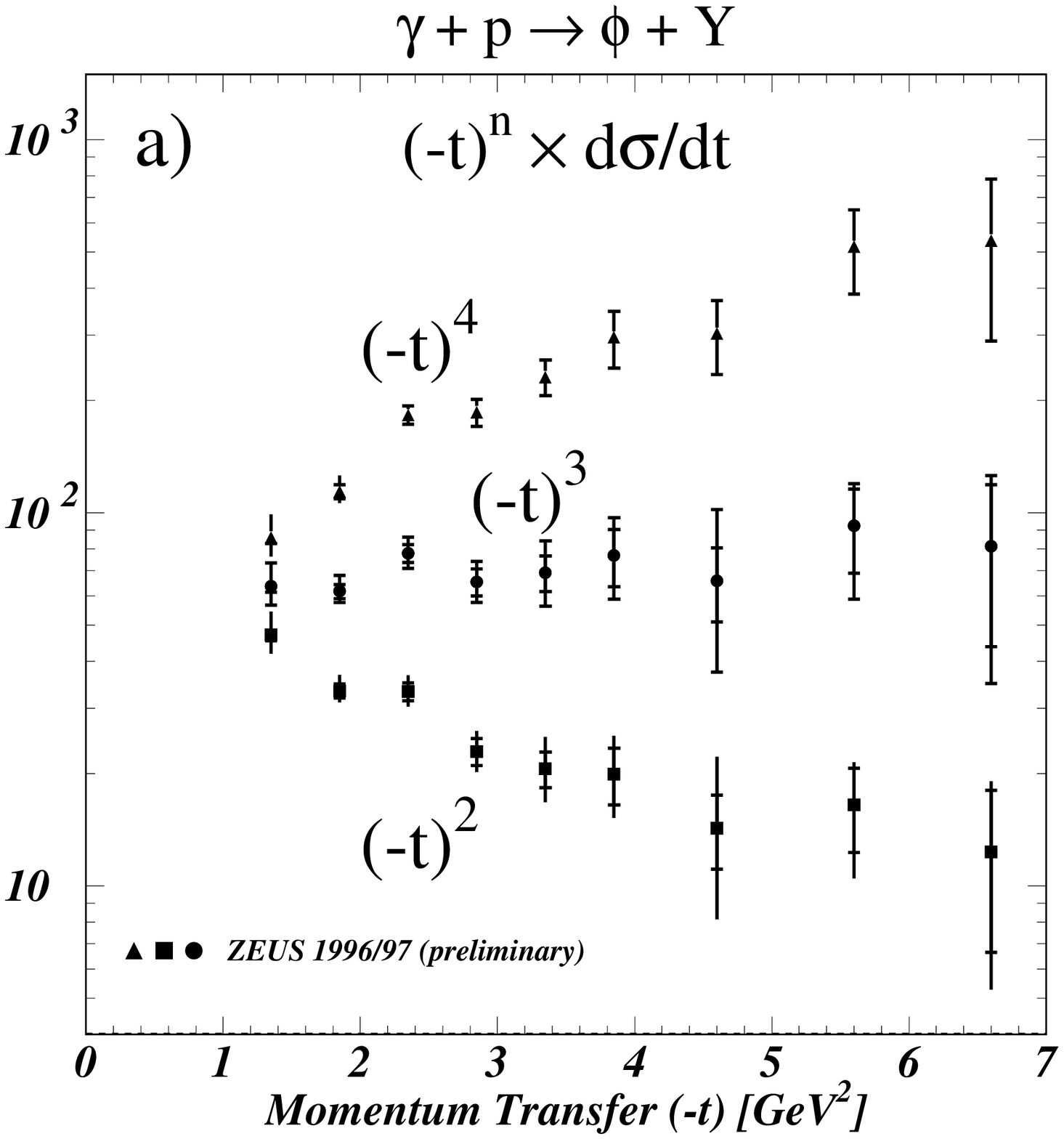, width=0.46\textwidth, bbllx=36, bblly=150, bburx=531, bbury=688, clip=}
\hfill
\epsfig{file=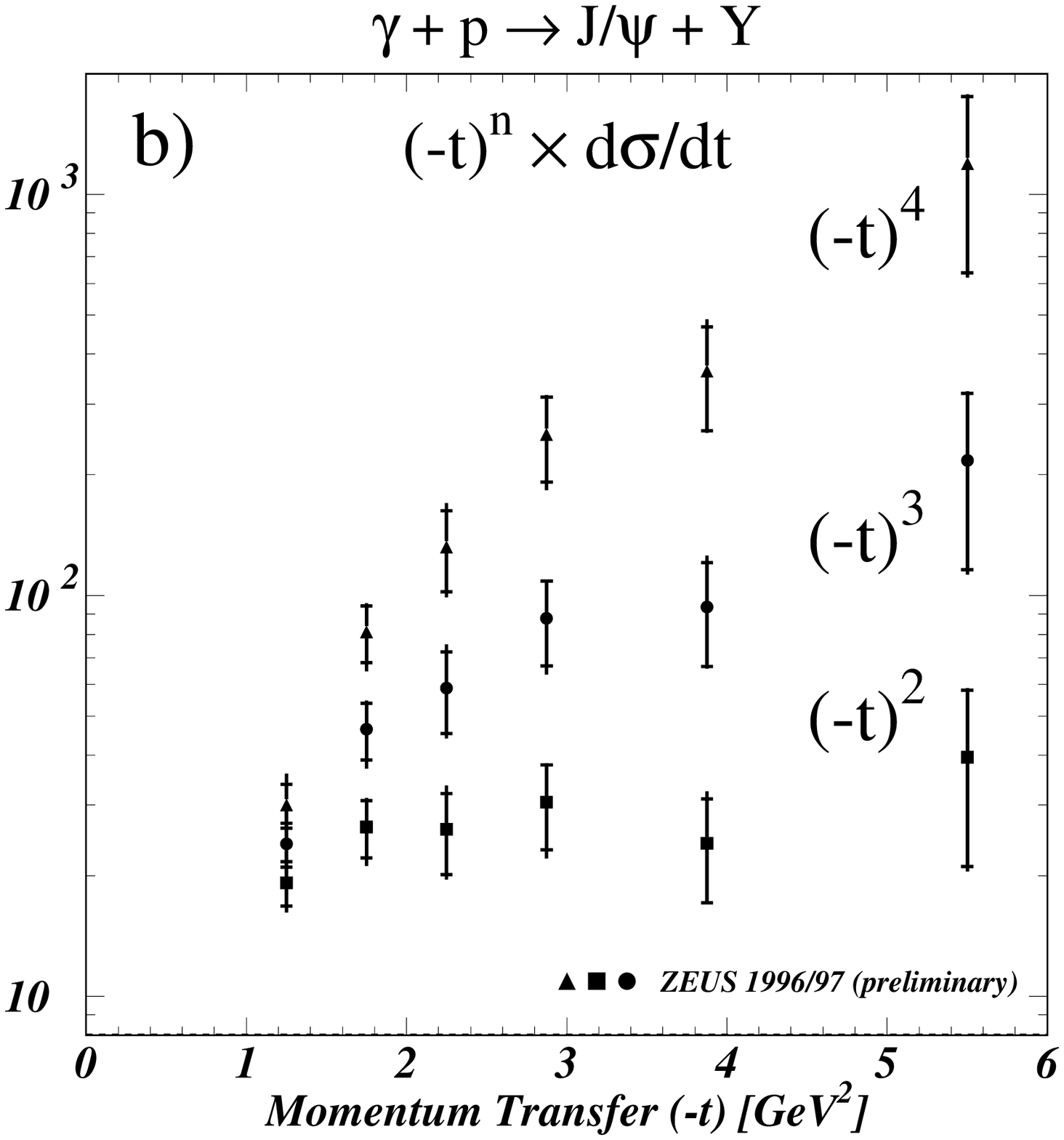, width=0.46\textwidth, bbllx=36, bblly=150, bburx=531, bbury=688, clip=}
\end{center}
\end{minipage}
\vspace*{-2mm}
\fcaption{
\label{fig:phijpsixsec}
Differential cross sections $\frac{\D\sigma}{\D t}$ 
for the process \mbox{$\gamma + p \rightarrow {\rm VM} + Y$} 
for (a)~{$\phi$} mesons and (b)~{\jpsi} mesons,  
multiplied by {(-t)}$^n$, where $n$=2, 3, 4.
}
\end{figure}

Such a scaling comparison is repeated for the differential cross sections
measured for $\phi$ and {\jpsi} photoproduction in 
Figs.~\ref{fig:phijpsixsec}a and~\ref{fig:phijpsixsec}b. The data exhibit
behavior consistent with a simple power law, with a power $n=3\;(2)$ for the
{$\phi$} ({\jpsi}).

\vspace*{1pt}\textlineskip   
\vspace*{-4mm}
\section{Summary}
\vspace*{-0.5pt}
\noindent
The principle features of the results can be summarized as
follows: 1) the {\rhoz} and $\phi$ differential cross sections may be
approximated by a simple power law $(-t)^{-n}$, where
$n\,{\simeq}\,3$.  Single powers of $n\,=\,2$ and 4 are clearly
excluded by the data over the $t$ range of the measurement. The
$t$ dependence for {\rhoz} photoproduction is somewhat steeper
than that for the $\phi$ in the region \mbox{$1.2<{\abst}<4~\gevsq$};
2) analysis of the decay angular distributions show that the {\rhoz}
and $\phi$ are produced with predominantly transverse polarization;
3) the ratio of $\phi$ to {\rhoz}
reaches the value of 2/9 consistent with
SU(3) flavor symmetry at high momentum transfer; 
4) the $t$~dependence observed for {\jpsi}
photoproduction shows a power $n\,{\simeq}\,2$ to describe the
dependence quite well, disfavoring higher powers. 
The flavor-symmetric value of 8/9 for
the $J\!{/}\!{\psi}$-to-${\rhoz}$ ratio is not observed in this region of momentum transfer.
%

\nonumsection{References}

\end{document}